\documentclass[]{ptptex}
\usepackage{wrapft}
\usepackage{graphicx}

\newcommand{\bra}[1]{\langle {#1} |}     
\newcommand{\rbra}[1]{( {#1} |}     
\newcommand{\rket}[1]{| {#1} )}     

\newcommand{\wtilde}[1]{\widetilde{#1}} 

\def\beq{\begin{eqnarray}}
\def\eeq{\end{eqnarray}}
\def\bsub{\begin{subequations}}
\def\esub{\end{subequations}}
\def\b{\begin{equation}}

\title{
A Role of the Quasiparticle in the Conservation of\\
the Fermion Number}
\subtitle{
An Example Illustrative of the Deformation of the Cooper Pair
}    

\author{
Yasuhiko {\sc Tsue},$^{1}$ 
Constan\c{c}a {\sc Provid\^encia},$^{2}$ 
Jo\~ao da {\sc Provid\^encia}$^{2}$ and 
Masatoshi {\sc Yamamura}$^{3}$  
}

\inst{
$^{1}$Physics Division, Faculty of Science, Kochi University, Kochi 780-8520, 
Japan\\
$^{2}$Departamento de F\'{i}sica, Universidade de Coimbra, 3004-516 Coimbra, 
Portugal\\
$^{3}$Department of Pure and Applied Physics, 
Faculty of Engineering Science,\\
Kansai University, Suita 564-8680, Japan
\\

}

\recdate{
\today
}

\abst{
The role of the ``quasiparticle" introduced in last paper by the present authors is discussed. 
It serves as a guide for the deformation of the Cooper-pair in the BCS-Bogoliubov approach. 
As an illustrative example, a certain case is treated. 
This case produces the $su(1,1)$-like behavior in the $su(2)$-algebraic many-fermion system. 
}

\begin{document}

\maketitle

In our last paper,\cite{1} which will be, hereafter, referred to as (I), 
we presented the quasiparticle in the conservation of the fermion number. 
Since this quasiparticle is not fermion operator, but the role is very similar 
to the conventional quasiparticle, we called it the ``quasiparticle." 
Through the medium of the ``quasiparticle," we showed that the BCS-Bogoliubov 
and the conventional algebraic approach are, in a certain sense, 
equivalent. 
In the process arriving at this conclusion, the following point played a basic role: 
the ``quasiparticle" is reduced to the conventional quasiparticle, if the $su(2)$-generators 
forming the Cooper-pair are replaced with $c$-numbers. 
The aim of this paper is to discuss one more role of the ``quasiparticle," 
namely, to present a possible guide for the deformation of the Cooper-pair 
in the BCS-Bogoliubov approach. 
In this paper, the interpretation of the symbols adopted (I) will be not repeated.

We will start in the condensed Cooper-pair state in the BCS-Bogoliubov approach shown in the 
relation (I$\cdot$2$\cdot$4): 
\beq\label{1}
\rket{\phi_B}=\frac{1}{\sqrt{{\Gamma}_B}}\exp\left(z{\wtilde S}_+\right)\rket{0}\ . \qquad
\left(z=\frac{v}{u}\ , \quad \Gamma_B=(1+|z|^2)^{2\Omega_0}\right)
\eeq
The state $\rket{\phi_B}$ is normalized and expanded in the form 
\beq
& &\rket{\phi_B}=\frac{1}{\sqrt{\Gamma_B}}\sum_{\sigma=0}^{2\Omega_0}
z^{\sigma}\sqrt{\frac{(2\Omega_0)!}{\sigma!(2\Omega_0-\sigma)!}}\rket{\sigma}\ , 
\label{2}\\
& &\rket{\sigma}=\sqrt{\frac{(2\Omega_0-\sigma)!}{\sigma!(2\Omega_0)!}}\left({\wtilde S}_+\right)^{\sigma}\rket{0}\ . 
\label{3}
\eeq
The state $\rket{\sigma}$ denotes the normalized state with $s=\Omega_0$ and 
$s_0=\sigma-\Omega_0$ and $\{\rket{\sigma}\}$ forms a complete set in the space 
with $s=\Omega_0$. 
The state (\ref{1}) obeys the condition 
\beq\label{4}
({\tilde c}_{\alpha}-zs_{\alpha}{\tilde c}_{\bar \alpha}^*)\rket{\phi_B}=0\ , \qquad
({\tilde c}_{\bar \alpha}+zs_{\alpha}{\tilde c}_{\alpha}^*)\rket{\phi_B}=0\ . 
\eeq
The condition (\ref{4}) enables us to define the quasiparticle and it leads to 
\bsub\label{5}
\beq
& &{\wtilde S}_-\rket{\phi_B}=z(\Omega_0-{\wtilde S}_0)\rket{\phi_B}\ , \qquad
(\Omega_0-{\wtilde S}_0=2\Omega_0-{\wtilde N}/2)
\label{5a}\\
& &(\Omega_0+{\wtilde S}_0)\rket{\phi_B}=z{\wtilde S}_+\rket{\phi_B}\ . 
\qquad
(\Omega_0+{\wtilde S}_0={\wtilde N}/2)
\label{5b}
\eeq
\esub
Here, ${\wtilde N}$ denotes the total fermion number operator. 
Operating ${\wtilde S}_+$ on the both sides of the relation (\ref{5a}) and using 
the relation (\ref{5b}), we obtain
\beq\label{6}
{\wtilde {\mib S}}^2\rket{\phi_B}=\Omega_0(\Omega_0+1)\rket{\phi_B}\ . 
\eeq
The operator ${\wtilde {\mib S}}^2$ denotes the Casimir operator (I$\cdot$2$\cdot$3). 
The relation (\ref{6}) is quite natural, because $\rket{\phi_B}$ is a superposition of the 
states (\ref{3}). 
The relation (\ref{5a}) can be re-formed to 
\beq\label{7}
& &(\Omega_0-{\wtilde S}_0+\epsilon)^{-1}{\wtilde S}_-\rket{\phi_B}=z(1-
\epsilon(\Omega_0-{\wtilde S}_0+\epsilon)^{-1})\rket{\phi_B}\ , 
\nonumber\\
& &\ \ 
\epsilon\ : \ {\rm infinitesimal\ positive\ parameter}.
\eeq
Here and hereafter, $\epsilon$ is used so as to be able to define the fractional operator. 
Then, the relation (\ref{4}) can be rewritten as 
\beq\label{8}
& &[{\tilde c}_{\alpha}-s_{\alpha}{\tilde c}_{\bar \alpha}^*(\Omega_0-{\wtilde S}_0+\epsilon)^{-1}{\wtilde S}_-]\rket{\phi_B}
=0\ , \quad
[{\tilde c}_{\bar \alpha}+s_{\alpha}{\tilde c}_{\alpha}^*(\Omega_0-{\wtilde S}_0+\epsilon)^{-1}{\wtilde S}_-]\rket{\phi_B}
=0\ .\nonumber\\
& &
\eeq
Starting in the state (\ref{1}), we arrived at the relation (\ref{8}), which does not 
depend formally on the parameter $z$. 
Of course, $\rket{\phi_B}$ is a special superposition of the states (\ref{3}). 
But, we should not forget that in the background of the relation (\ref{8}), 
there exists the relation (\ref{4}).

Next, we take up the ``quasiparticle" obeying the condition (I$\cdot$3$\cdot$2):
\bsub\label{9}
\beq
& &{\tilde d}_{\alpha +}=\frac{1}{2\Omega_0}
[{\tilde c}_{\alpha}(\Omega_0-{\wtilde S}_0)-{\wtilde S}_-s_{\alpha}{\tilde c}_{\bar \alpha}^*]\ , \quad
{\tilde d}_{{\bar \alpha} +}=\frac{1}{2\Omega_0}
[{\tilde c}_{\bar \alpha}(\Omega_0-{\wtilde S}_0)+{\wtilde S}_-s_{\alpha}{\tilde c}_{\alpha}^*]\ , 
\label{9a}\\
& &{\tilde d}_{\alpha -}=-\frac{1}{2\Omega_0}
[{\wtilde S}_+{\tilde c}_{\alpha}-s_{\alpha}{\tilde c}_{\bar \alpha}^*(\Omega_0+{\wtilde S}_0)]\ , \quad
{\tilde d}_{{\bar \alpha} -}=-\frac{1}{2\Omega_0}
[{\wtilde S}_+{\tilde c}_{\bar \alpha}+s_{\alpha}{\tilde c}_{\alpha}^*(\Omega_0+{\wtilde S}_0)]\ . \ \ \ \ 
\label{9b}
\eeq
\esub
It was shown in (I) that, for any value of $\sigma$, the ``quasiparticle" satisfies 
\beq\label{10}
{\tilde d}_{\alpha +}\rket{\sigma}={\tilde d}_{{\bar \alpha} +}\rket{\sigma}
={\tilde d}_{\alpha -}\rket{\sigma}={\tilde d}_{{\bar \alpha} -}\rket{\sigma}=0 \ .
\eeq
Therefore, for any state $\rket{\phi}$ which is a superposition of the states (\ref{3}) ($\rket{\sigma}$), 
we have 
\bsub\label{11}
\beq
& &{\tilde d}_{\alpha +}\rket{\phi}={\tilde d}_{{\bar \alpha}+}\rket{\phi}=0\ , 
\label{11a}\\
& &{\tilde d}_{\alpha -}\rket{\phi}={\tilde d}_{{\bar \alpha}-}\rket{\phi}=0\ . 
\label{11b}
\eeq
\esub
Of course, $\rket{\phi_B}$ is an example of $\rket{\phi}$. 
After straightforward calculation, we can re-form the operators ${\tilde d}_{\alpha +}$ and 
${\tilde d}_{{\bar \alpha}+}$ in the form 
\beq\label{12}
& &{\tilde d}_{\alpha +}=\frac{1}{2\Omega_0}
\left(\Omega_0-{\wtilde S}_0+\frac{1}{2}\right)
\left[{\tilde c}_{\alpha}-s_{\alpha}{\tilde c}_{{\bar \alpha}}^*(\Omega_0-{\wtilde S}_0+\epsilon)^{-1}{\wtilde S}_-\right]\ , 
\nonumber\\
& &{\tilde d}_{{\bar \alpha} +}=\frac{1}{2\Omega_0}
\left(\Omega_0-{\wtilde S}_0+\frac{1}{2}\right)
\left[{\tilde c}_{\bar \alpha}+s_{\alpha}{\tilde c}_{{\alpha}}^*(\Omega_0-{\wtilde S}_0+\epsilon)^{-1}{\wtilde S}_-\right]\ . 
\eeq
The relation (\ref{8}) was derived for the state $\rket{\phi_B}$. 
But, the relations (\ref{11a}) and (\ref{12}) suggest us that the relation (\ref{8}) holds in an arbitrary state 
$\rket{\phi}$ including $\rket{\phi_B}$: 
\beq\label{13}
& &[{\tilde c}_{\alpha}-s_{\alpha}{\tilde c}_{{\bar \alpha}}^*(\Omega_0-{\wtilde S}_0+\epsilon)^{-1}{\wtilde S}_-]\rket{\phi}=0\ , \quad
[{\tilde c}_{\bar \alpha}+s_{\alpha}{\tilde c}_{{\alpha}}^*(\Omega_0-{\wtilde S}_0+\epsilon)^{-1}{\wtilde S}_-]\rket{\phi}=0\ .\nonumber\\
& & 
\eeq
Next, we consider the case (\ref{9b}), which is rewritten as 
\beq
& &{\tilde d}_{\alpha -}=-\frac{1}{2\Omega_0}{\wtilde S}_+
[{\tilde c}_{\alpha}-s_{\alpha}{\tilde c}_{\bar \alpha}^*(\Omega_0-{\wtilde S}_0+\epsilon)^{-1}
{\tilde \rho}{\wtilde S}_-]\ , 
\nonumber\\
& &{\tilde d}_{{\bar \alpha} -}=-\frac{1}{2\Omega_0}{\wtilde S}_+
[{\tilde c}_{\bar \alpha}+s_{\alpha}{\tilde c}_{\alpha}^*(\Omega_0-{\wtilde S}_0+\epsilon)^{-1}
{\tilde \rho}{\wtilde S}_-]\ , 
\label{14}\\
& &{\tilde \rho}=(\Omega_0-{\wtilde S}_0)(\Omega_0+{\wtilde S}_0+1)
\left(({\wtilde S}-{\wtilde S}_0)({\wtilde S}+{\wtilde S}_0+1)+\epsilon\right)^{-1}\ . 
\label{15}
\eeq
Here, ${\wtilde S}$ denotes the operator expressing the magnitude of the $S$-spin: 
\beq\label{16}
({\wtilde S}-{\wtilde S}_0)({\wtilde S}+{\wtilde S}_0+1)={\wtilde {\mib S}}^2
-{\wtilde S}_0({\wtilde S}_0+1)
={\wtilde S}_-{\wtilde S}_+\ . 
\eeq
Since ${\tilde \rho}\rket{\phi}=\rket{\phi}$, the relation (\ref{11b}) reduces to 
the relation (\ref{13}). 
Then, we can understand that the relation (\ref{13}) is regarded as the condition which 
governs any state with $s=\Omega_0$. 
Therefore, only within the framework of the condition (\ref{13}), it is impossible to determine 
the explicit form of $\rket{\phi}$. 
For the determination of $\rket{\phi}$, some condition should be added. 
In the case of the BCS-Bogoliubov approach, 
the following is adopted: 
\beq\label{17}
{\wtilde S}_-\rket{\phi}=z(\Omega_0-{\wtilde S}_0)\rket{\phi}\ . 
\eeq
Through the condition (\ref{17}), $\rket{\phi}$ becomes a specific superposition of the states 
(\ref{3}), which is shown in the form (\ref{2}). 
The above is our guide to the deformation of the Cooper-pair in the BCS-Bogoliubov approach.

As an example of the deformation, in this paper, we will investigate the following case:
\beq\label{18}
{\wtilde S}_-\rket{\phi}
=z\sqrt{(\Omega_0-{\wtilde S}_0)(\Omega_0+{\wtilde S}_0+1)}\rket{\phi}\ .
\eeq
The conditions (\ref{17}) and (\ref{18}) should be compared with one another. 
The reasons why we investigate the case (\ref{18}) are as follows: 
(1) The results are expressed in rather simple form and (2) in contrast with 
the $su(2)$-algebraic structure based on the condition (\ref{17}), the condition 
(\ref{18}) presents the $su(1,1)$-like behavior. 
Of course, these will be shown later. 
Substituting the relation (\ref{18}) into the relation (\ref{13}), 
we have 
\beq\label{19}
& &\left(
{\tilde c}_{\alpha}-zs_{\alpha}{\tilde c}_{\bar \alpha}^*
\sqrt{(\Omega_0-{\wtilde S}_0+\epsilon)^{-1}(\Omega_0+{\wtilde S}_0+1)}\right)\rket{\phi}=0\ , 
\nonumber\\
& &\left(
{\tilde c}_{\bar \alpha}+zs_{\alpha}{\tilde c}_{\alpha}^*
\sqrt{(\Omega_0-{\wtilde S}_0+\epsilon)^{-1}(\Omega_0+{\wtilde S}_0+1)}\right)\rket{\phi}=0\ . 
\eeq
The relation (\ref{19}) leads to two relations. 
One is the relation (\ref{18}) itself and the other is as follows: 
\beq\label{20}
(\Omega_0+{\wtilde S}_0)\rket{\phi}=
z{\wtilde S}_+\sqrt{(\Omega_0-{\wtilde S}_0+\epsilon)^{-1}({\Omega_0}+{\wtilde S}_0+1)}\rket{\phi}\ . \ \
(\Omega_0+{\wtilde S}_0={\wtilde N}/2)
\eeq
The relations (\ref{18}) and (\ref{20}) give the following: 
\beq
& &\rbra{\phi}\Omega_0+{\wtilde S}_0\rket{\phi}
=\frac{|z|^2}{1-|z|^2}\left[1-(2\Omega_0+1)|(2\Omega_0\rket{\phi}|^2\right]\ , 
\label{21}\\
& &
\rbra{\phi}{\wtilde S}_+\sqrt{(\Omega_0-{\wtilde S}_0+\epsilon)^{-1}(\Omega_0+{\wtilde S}_0+1)}\rket{\phi}
=\frac{z^*}{1-|z|^2}\left[1-(2\Omega_0+1)|(2\Omega_0\rket{\phi}|^2\right]\ . \nonumber\\
& &
\label{22}
\eeq
The case with $|z|^2=1$ is excluded from the relations (\ref{21}) and (\ref{22}). 
In order to obtain the definite expressions, we must determine the quantity 
$(2\Omega_0\rket{\phi}$.

First, we note that the condition (\ref{18}) gives
\beq\label{23}
\rbra{\alpha-1}{\wtilde S}_-\rket{\phi}=z\sqrt{(2\Omega_0-\alpha+1)\alpha}(\alpha-1\rket{\phi}\ . 
\qquad (\alpha=1,\ 2, \cdots ,\ 2\Omega_0)
\eeq
Using the formula ${\wtilde S}_+\rket{\alpha-1}=\sqrt{(2\Omega_0-\alpha+1)\alpha}\rket{\alpha}$, we have 
\bsub\label{24}
\beq
(\alpha\rket{\phi}=z(\alpha-1\rket{\phi}\ . 
\label{24a}
\eeq
The recursion formula (\ref{24a}) leads to 
\beq
(\alpha\rket{\phi}=z^{\alpha}(0\rket{\phi}\ . 
\label{24b}
\eeq
\esub
Since $\rket{\phi}$ should be normalized, we have 
\beq\label{25}
& &(\phi\rket{\phi}=\sum_{\alpha=0}^{2\Omega_0}
|(\alpha\rket{\phi}|^2=\sum_{\alpha=0}^{2\Omega_0}
(|z|^2)^{\alpha}\cdot|(0\rket{\phi}|^2=1\ , \nonumber\\
{\rm i.e.,}\quad
& & (0\rket{\phi}=\frac{1}{\sqrt{\Gamma}}\ , \qquad
\Gamma=\sum_{\alpha=0}^{2\Omega_0}(|z|^2)^{\alpha}
=\frac{1-(|z|^2)^{2\Omega_0+1}}{1-|z|^2}\ . \qquad
(|z|^2\neq 1)
\eeq
The relations (\ref{24b}) and (\ref{25}) give
\beq\label{26}
|(2\Omega_0\rket{\phi}|^2=\frac{(|z|^2)^{2\Omega_0}}{\Gamma}\ . 
\eeq
Thus, we can derive the explicit expressions for the relations (\ref{21}) and (\ref{23}) in the form 
\bsub\label{27}
\beq
& &\rbra{\phi}\Omega_0+{\wtilde S}_0\rket{\phi}=\frac{1}{2}N=\frac{|z|^2}{1-|z|^2}-
(2\Omega_0+1)\cdot\frac{(|z|^2)^{2\Omega_0+1}}{1-(|z|^2)^{2\Omega_0+1}}\ , 
\label{27a}\\
& &\rbra{\phi}{\wtilde S}_+\sqrt{(\Omega_0-{\wtilde S}_0+\epsilon)^{-1}(\Omega_0+{\wtilde S}_0+1)}\rket{\phi}
={\cal T}_+\nonumber\\
& &\qquad\qquad\qquad
=z^*\left(\frac{1}{1-|z|^2}-(2\Omega_0+1)\cdot\frac{(|z|^2)^{2\Omega_0}}{1-(|z|^2)^{2\Omega_0+1}}\right)\ . 
\label{27b}
\eeq
\esub
Here, $N$ denotes the average value $\rbra{\phi}{\wtilde N}\rket{\phi}$ and later the reason 
why we used the notation ${\cal T}_+$ will be mentioned. 
The above two are for $|z|^2\neq 1$ and the case with $|z|^2=1$ is given as 
\beq\label{28}
\frac{1}{2}N=\Omega_0\ , \qquad
{\cal T}_+=\Omega_0 e^{i\theta}\ . \qquad
(z^*=e^{i\theta})
\eeq
Although the form (\ref{27}) was derived under the condition $|z|^2\neq 1$, the limit value at 
$|z|^2\rightarrow 1$ coincides with the value (\ref{28}). 
Namely, they are continuous at $|z|^2=1$. 
On the occasion of closing this paragraph, we mention two comments: 
(1) The explicit expression of $\rket{\phi}$ satisfying the relations (\ref{24}) and (\ref{25}) is given by 
\beq\label{29}
\rket{\phi}
=\frac{1}{\sqrt{\Gamma}}\exp\left(z{\wtilde S}_+\sqrt{(\Omega_0-{\wtilde S}_0+\epsilon)^{-1}
(\Omega_0+{\wtilde S}_0+1)}\right)\rket{0}
=\frac{1}{\sqrt{\Gamma}}\sum_{\sigma=0}^{2\Omega_0}z^{\sigma}\rket{\sigma}\ . 
\eeq
Of course, $\Gamma$ for $|z|^2\neq 1$ is given in the relation (\ref{25}) and the case with $|z|^2=1$ 
is given as $\Gamma=2\Omega_0+1$. 
(2) With the use of the state (\ref{29}), we have the following expression for $N/2$ and ${\cal T}_+$ for any 
value of $z$: 
\beq\label{30}
\frac{1}{2}N=\frac{1}{\Gamma}\sum_{\sigma=1}^{2\Omega_0}\sigma (|z|^2)^{\sigma}\ , \qquad
{\cal T}_+=\frac{z^*}{\Gamma}\sum_{\sigma=1}^{2\Omega_0}\sigma(|z|^2)^{\sigma-1}\ . 
\eeq

The quantity $N/2$ is an increasing function of $|z|^2$. 
Then, hereafter, we denote it as $N(|z|^2)/2$. 
At the three points, we have $N(0)/2=0$, $N(1)/2=\Omega_0$ and $N(\infty)/2=2\Omega_0$. 
Further, the following relation is noticeable: 
\beq\label{31}
\frac{1}{2}N(|z|^2)+\frac{1}{2}N(1/|z|^2)=2\Omega_0\ , \quad
{\rm i.e.,}\quad 
\frac{1}{2}N(1/|z|^2)=2\Omega_0-\frac{1}{2}N(|z|^2)\ . 
\eeq
Through the straightforward calculation, the relation (\ref{31}) can be derived. 
The relation (\ref{31}) tells us that it may be enough to investigate only the case 
with $(0\leq |z|^2 \leq 1$, $0\leq N/2 \leq \Omega_0$). 
By replacing $|z|^2$ with $1/|z|^2$, we can treat the case 
($1\leq |z|^2 <\infty$, $\Omega_0 \leq N/2 \leq 2\Omega_0$). 
Next, we consider the case (\ref{27b}). 
Combining with the expression (\ref{27a}), the relation (\ref{27b}) can be rewritten as 
\beq\label{32}
{\cal T}_+=e^{i\theta}\cdot {\cal T}(|z|^2)\ , \qquad
{\cal T}(|z|^2)=\frac{1}{\sqrt{|z|^2}}\cdot\frac{1}{2}N(|z|^2)\ . \quad 
(z^*=e^{i\theta}|z|)
\eeq
Therefore, we have 
\beq\label{33}
{\cal T}_+(1/|z|^2)=\sqrt{|z|^2}\cdot\frac{1}{2}N(1/|z|^2)=\sqrt{|z|^2}\left(2\Omega_0-\frac{1}{2}N(|z|^2)\right)\ .
\eeq
We can learn that for the quantity ${\cal T}_+$, it is enough to treat only the case 
($0\leq |z|^2 \leq 1$, $0\leq N/2\leq 2\Omega_0$). 
However, $\rbra{\phi}{\wtilde S}_+\rket{\phi}$ cannot be expressed in a simple form: 
\beq\label{34}
\rbra{\phi}{\wtilde S}_+\rket{\phi}=\frac{z^*}{\Gamma}\sum_{\sigma=0}^{2\Omega_0-1}
(|z|^2)^{\sigma}\sqrt{(\sigma+1)(2\Omega_0-\sigma)}\ .
\eeq
If $|z|^2/(1-|z|^2) \gg (2\Omega_0+1)\cdot (|z|^2)^{2\Omega_0+1}/(1-(|z|^2)^{2\Omega_0+1})$ and 
$N/2 \ll \Omega_0$ in the relation (\ref{27a}), the expression (\ref{34}) can be approximated to 
\beq\label{35}
\rbra{\phi}{\wtilde S}_+\rket{\phi}\approx e^{i\theta}\cdot 2\Omega_0
\sqrt{\frac{N}{4\Omega_0}\left(1-\frac{N}{4\Omega_0}\right)}\ . 
\eeq
The above is of the same form as that in the BCS-Bogoliubov approach. 
Further, the following should be noted: 
Usually, as the Hamiltonian of the pairing model, the form ${\wtilde H}=-G{\wtilde S}_+{\wtilde S}_-$ 
($G$ : coupling constant, positive) is adopted. 
In the case $N/2=\Omega_0$, the expectation values 
of ${\wtilde H}$, $E$, are easily calculated in the form 
$E_{\rm exact}=-G\Omega_0(\Omega_0+1)$, 
$E_{\rm BCS}=E_{\rm exact}+2G\Omega_0$ and 
$E_{\rm present}=E_{\rm exact}+(G/3)\cdot \Omega_0(\Omega_0+1)$. 
Since usually $\Omega_0$ is sufficiently large and $0<2G\Omega_0\ll (G/3)\cdot\Omega_0(\Omega_0+1)$, 
we have $E_{\rm exact}<E_{\rm BCS}\ll E_{\rm present}$. 
This inequality tells that for the description of superconductivity, the BCS-Bogoliubov approach is 
superior to the present. 
Then, we have a question: 
For what purpose, the present approach must be investigated ?
At the ending part of this paper, we will give an answer to this question.

From a viewpoint slightly different from the above, we will investigate the present problem. 
First, we define the following operators: 
\bsub
\beq
& &{\wtilde {\cal T}}_+={\wtilde S}_+\cdot\sqrt{(\Omega_0-{\wtilde S}_0+\epsilon)^{-1}
(\Omega_0+{\wtilde S}_0+1)}\ , 
\label{35a}\\
& &{\wtilde {\cal T}}_-=\sqrt{(\Omega_0-{\wtilde S}_0+\epsilon)^{-1}
(\Omega_0+{\wtilde S}_0+1)}\cdot{\wtilde S}_-\ , 
\label{35b}\\
& &{\wtilde {\cal T}}_0={\wtilde S}_0+\Omega_0+\frac{1}{2}\  
\label{35c}
\eeq
\esub
The state (\ref{29}) can be expressed in the form 
\beq\label{36}
\rket{\phi}=\frac{1}{\sqrt{\Gamma}}\exp\left(z{\wtilde {\cal T}}_+\right)\rket{0}\ . 
\eeq
Comparison of $\rket{\phi}$ with $\rket{\phi_B}$ suggests that the set of the $su(2)$-generators 
$({\wtilde S}_{\pm,0})$ is deformed to $({\wtilde {\cal T}}_{\pm,0})$: 
\beq\label{37}
{\wtilde S}_{\pm,0} \stackrel{\rm (deformed)}{\longrightarrow} {\wtilde {\cal T}}_{\pm,0}\ . 
\eeq
In the space with $s=\Omega_0$, the following commutation relation is derived: 
\beq\label{38}
& &[\ {\wtilde {\cal T}}_+\ , \ {\wtilde {\cal T}}_-\ ]=-2{\wtilde {\cal T}}_0+(2\Omega_0+1)^2
\rket{2\Omega_0}\rbra{2\Omega_0}\ , \nonumber\\
& &[\ {\wtilde {\cal T}}_0\ , \ {\wtilde {\cal T}}_{\pm}\ ]=\pm{\wtilde {\cal T}}_{\pm}\ .
\eeq
Further, we have 
\beq\label{39}
{\wtilde {\mib {\cal T}}}^2&=&{\wtilde {\cal T}}_0^2-\frac{1}{2}
\left({\wtilde {\cal T}}_+{\wtilde {\cal T}}_-+{\wtilde {\cal T}}_-{\wtilde {\cal T}}_+\right)
\nonumber\\
&=&\frac{1}{2}\left(\frac{1}{2}-1\right)+\frac{1}{2}(2\Omega_0+1)^2\rket{2\Omega_0}\rbra{2\Omega_0}\ . 
\eeq
It may be interesting to see that if the effects from the term $\rket{2\Omega_0}\rbra{2\Omega_0}$ 
can be neglected, the set $({\wtilde {\cal T}}_{\pm,0})$ forms the $su(1,1)$-algebra and 
${\wtilde {\mib {\cal T}}}^2$ is regarded as the Casimir operator, the eigenvalue of which is given by 
$t(t-1)$ with the magnitude of the $su(1,1)$-spin $t=1/2$. 
This feature was discussed by the present authors with Kuriyama by the name of the pseudo $su(1,1)$-algebra 
in many-boson system.\cite{2} 
Certainly, as is shown in Fig.1, the range $0\leq |z|^2 \ll 1$ supports this fact. 
In the case of the $su(1,1)$-algebra, at the limit $|z|^2\rightarrow 1$, $N/2\rightarrow \infty$ and we can see that the $su(1,1)$-algebraic 
behavior breaks down. 
The $su(2)$ and the $su(1,1)$-algebra are compact and non-compact, respectively. 
If $\Omega_0$ is sufficiently large as is shown in Fig.2 (b), there exist many-states 
$(\sigma=0,\ 1,\ 2,\cdots ,\ 2\Omega_0-1,\ 2\Omega_0)$ and it may be permissible to neglect the effect 
of the term $\rket{2\Omega_0}\rbra{2\Omega_0}$. 
Therefore, the existence of $\rket{2\Omega_0}\rbra{2\Omega_0}$ supports the compactness of the set 
$({\wtilde {\cal T}}_{\pm,0})$ deformed from the set $({\wtilde S}_{\pm,0})$. 
But, the behavior of $({\wtilde {\cal T}}_{\pm,0})$ seems to be very different from that shown 
by the state $\rket{\phi_B}$.

\begin{figure}[t]
\begin{center}
\includegraphics[height=4.8cm]{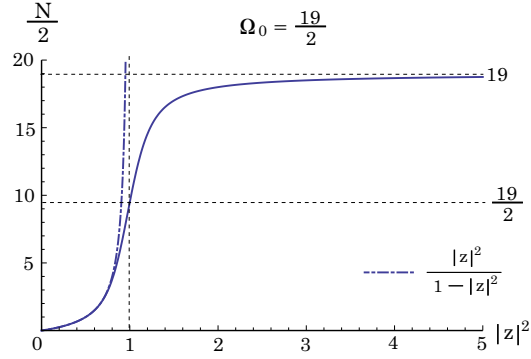}
\caption{
The figure shows $N/2$ as a function of $|z|^2$ for the case $\Omega_0=19/2$. 
The dash-dotted curves represents the result based on the $su(1,1)$-algebra. 
}
\label{fig:1}
\end{center}
\end{figure}

\begin{figure}[t]
\begin{center}
\includegraphics[height=4.8cm]{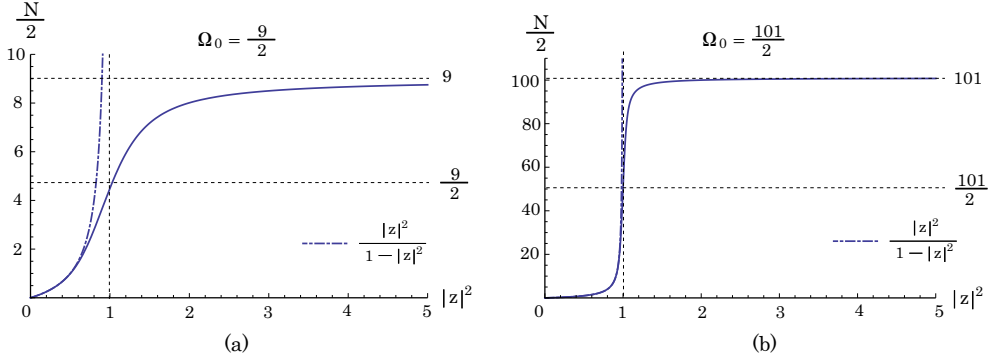}
\caption{
The figure shows $N/2$ as a function of $|z|^2$ for the case with (a) small $\Omega_0$ ($\Omega_0=9/2$) and 
(b) large $\Omega_0$ ($\Omega_0=101/2$), respectively. 
The dash-dotted curves represents the result based on the $su(1,1)$-algebra. 
}
\label{fig:2}
\end{center}
\end{figure}

We can see the above feature in the form of $\rbra{\phi}{\wtilde {\cal T}}_{\pm,0}\rket{\phi}$, 
which can be calculated as 
\bsub\label{40}
\beq
& &\bra{\phi}{\wtilde {\cal T}}_+\rket{\phi}=y^*\sqrt{1+y^*y-(2\Omega_0+1)\frac{(|z|^2)^{2\Omega_0}(1-|z|^2)}{\Gamma}}\ , 
\label{40a}\\
& &\bra{\phi}{\wtilde {\cal T}}_-\rket{\phi}=y\sqrt{1+y^*y-(2\Omega_0+1)\frac{(|z|^2)^{2\Omega_0}(1-|z|^2)}{\Gamma}}\ , 
\label{40b}\\
& &\bra{\phi}{\wtilde {\cal T}}_0\rket{\phi}=y^*y+\frac{1}{2}\ . 
\label{40c}
\eeq
\esub
Here, $y$ is defined as 
\beq\label{41}
y=z\sqrt{\frac{1}{1-|z|^2}-(2\Omega_0+1)\frac{(|z|^2)^{2\Omega_0}}{1-(|z|^2)^{2\Omega_0+1}}}\ . 
\eeq
The expression (\ref{40a}) is equivalent to ${\cal T}_+$ given in the relation (\ref{27b}), which 
explains the reason why we used the notation ${\cal T}_+$. 
It should be noted that $y^*y$ is reduced to 
\beq\label{42}
y^*y=\frac{1}{2}N\ .
\eeq
If the second term contained in the square root of the relation (\ref{40}) is neglected, 
the expression (\ref{40}) is reduced to the classical counterpart of the $su(1,1)$-algebra presented by the present authors 
with Kuriyama.\cite{2}

The above two examples suggest us to give a possible extension. 
We set up the following form: 
\beq\label{43}
{\wtilde S}_-\rket{\phi}=z(\Omega_0-{\wtilde S}_0)f(\Omega_0+{\wtilde S}_0; \Omega_0,\epsilon)\rket{\phi}\ .
\eeq
If $f(\Omega_0+{\wtilde S}_0;\Omega_0,\epsilon)=1$ and 
$\sqrt{(\Omega_0-{\wtilde S}_0+\epsilon)^{-1}(\Omega_0+{\wtilde S}_0+1)}$, 
the form (\ref{43}) reduces to the relations (\ref{17}) and (\ref{18}), respectively. 
Substituting the form (\ref{43}) into the relation (\ref{13}), we have 
\beq\label{44}
[{\tilde c}_{\alpha}-zs_{\alpha}{\tilde c}_{\bar \alpha}^*f(\Omega_0+{\wtilde S}_0;\Omega_0,\epsilon)]\rket{\phi}
=0\ , \quad
[{\tilde c}_{\bar \alpha}+zs_{\alpha}{\tilde c}_{\alpha}^*f(\Omega_0+{\wtilde S}_0;\Omega_0,\epsilon)]\rket{\phi}
=0\ .\quad\ \ 
\eeq
Then, the following relation is derived: 
\beq\label{45}
(\Omega_0+{\wtilde S}_0)\rket{\phi}=z{\wtilde S}_+f(\Omega_0+{\wtilde S}_0;\Omega_0,\epsilon)\rket{\phi}\ .
\eeq
The relations (\ref{43}) and (\ref{45}) give 
\beq\label{46}
& &\rbra{\phi}\Omega_0+{\wtilde S}_0\rket{\phi}=|z|^2
\rbra{\phi}(\Omega_0-{\wtilde S}_0)f(\Omega_0+{\wtilde S}_0;\Omega_0,\epsilon)^2\rket{\phi}\ , \nonumber\\
& &\rbra{\phi}{\wtilde S}_+f(\Omega_0+{\wtilde S}_0;\Omega_0,\epsilon)\rket{\phi}=
z^*\rbra{\phi}(\Omega_0-{\wtilde S}_0)f(\Omega_0+{\wtilde S}_0;\Omega_0,\epsilon)^2\rket{\phi}
\nonumber\\
& &\qquad\qquad\qquad\qquad\qquad\quad\ 
=\frac{z^*}{|z|^2}\cdot\rbra{\phi}\Omega_0+{\wtilde S}_0\rket{\phi}\ . 
\eeq
The quantity $(\sigma\rket{\phi}$ is obtained in the form 
\beq\label{47}
& &(\sigma\rket{\phi}=z^{\sigma}
\sqrt{\frac{(2\Omega_0)!}{\sigma!(2\Omega_0-\sigma)!}}\prod_{\rho=0}^{\sigma-1}f(\rho;\Omega_0,\epsilon)\ , \nonumber\\
& &\sum_{\sigma}|(\sigma\rket{\phi}|^2=\sum_{\sigma}(|z|^2)^{\sigma}\frac{(2\Omega_0)!}{\sigma!(2\Omega_0-\sigma)!}
\prod_{\rho=0}^{\sigma-1}f(\rho;\Omega_0,\epsilon)^2\cdot |(0\rket{\phi}|^2=1\ . 
\eeq
The state $\rket{\phi}$ is expressed explicitly as 
\bsub\label{48}
\beq
\rket{\phi}&=&
\frac{1}{\sqrt{\Gamma}}\exp\left(z{\wtilde S}_+f(\Omega_0+{\wtilde S}_0;\Omega_0,\epsilon)\right)\rket{0}\nonumber\\
&=&\frac{1}{\sqrt{\Gamma}}\sum_{\sigma=0}^{2\Omega_0}\sqrt{\frac{(2\Omega_0)!}{\sigma!(2\Omega_0-\sigma)!}}\prod_{\rho=0}^{\sigma-1}
f(\rho;\Omega_0,\epsilon)\rket{\sigma}\ , 
\label{48a}\\
\Gamma&=&
\sum_{\sigma=0}^{2\Omega_0}(|z|^2)^{\sigma}\frac{(2\Omega_0)!}{\sigma!(2\Omega_0-\sigma)!}\prod_{\rho=0}^{\sigma-1}
f(\rho;\Omega_0,\epsilon)^2\ . 
\label{48b}
\eeq
\esub
The above is the extension from the form discussed concretely in this paper.

In the introductory part, the aim of this paper was mentioned. 
We did not intend to give any application to concrete problems. 
As a final remark, let us note down a possibility of the application of the basic idea. 
Following an idea discussed by Celeghini et al.,\cite{3} 
the present authors with Kuriyama\cite{2} investigated 
various features of the $su(1,1)$-algebra based on the Schwinger boson representation.\cite{4}
With the aid of this investigation, the ``damped and amplified oscillator" can be described quantum mechanically in the 
conservation form. 
Further, in the framework of the $su(2)$-Schwinger boson representation,\cite{4} 
the pseudo $su(1,1)$-algebra we called is defined and by using this form, we can describe 
emission and absorption of the bosons. 
Through this description, various interesting aspects of thermal property and dissipation for many-boson 
system are investigated. 
Its theoretical background is the thermo field 
dynamics formalism developed by Umezawa et al.\cite{5} 
The above mention on the many-boson system suggests that the idea developed in this paper can help 
to describe thermal property and dissipation for many-fermion system. 
For this aim, the form presented in this paper may be necessary to extend to the case 
$(t>1/2,\ s<\Omega_0)$ and also, the $R$-spin may be useful. 
This spin was introduced in the papers which preceded to the present one.\cite{6}
The Cooper-pair plays a decisive role for describing the superconductivity. 
On the other hand, as was already mentioned, its possible deformation may be 
helpful for the understanding of thermal property and dissipation for many-fermion system. 
Clearly, this approach has no connection with the description of the superconductivity directly. 
This means that the present deformation enriches the role of the $su(2)$-algebraic model 
for many-fermion system. 
Of course, in this case, ${\wtilde S}_{\pm}$ cannot be regarded as Cooper-pair operators. 
The above is the answer to the question we have mentioned and the concrete treatment is our 
future problem.

\vspace{1cm}
\section*{Acknowledgement}

One of the authors (Y.T.) 
is partially supported by the Grants-in-Aid of the Scientific Research 
(No.23540311) from the Ministry of Education, Culture, Sports, Science and 
Technology in Japan.


\end{document}